\documentclass[11pt]{article}
\usepackage{hyperref}
\pdfoutput=1
\usepackage{natbib}
\begin{document}
\title{Marching of Freely Falling Plates}
\author{Hui Wan, Haibo Dong, Zachary Gaston, Zongxian Liang \\
        \vspace{6pt} Department of Mechanical and Materials Engineering,
	Wright State University\\
	\vspace{6pt} Dayton, OH 45435, USA}
\maketitle

\begin{abstract}
``Marching of freely falling plates'' is a fluid dynamics video submitted to the Gallery of Fluid Motion in APS-DFD 2011 at Baltimore Maryland.
The problem of a freely falling plate is of interest in both fluid mechanics and nonlinear dynamics.
The trajectory of the plate can be regular (Willmarth et al., 1964) or chaotic (Aref and Jones, 1993). 
As long as Reynolds number is high enough, regular flutter and tumble motion can be obtained for plates with small and large Froude numbers respectively.
Belmonte et al. (1998) conducted experimental study on thin flat strips falling in a vertical cell. They categorized the Froude number at which the transition from flutter to tumble occurs.
Andersen et al. (2005) analyzed the transitions between fluttering and tumbling using vorticity-stream function formulation and ODE dynamic equations based on quasi-steady models. They also found that the fluid circulation is mainly generated by the plate rotation and its angular velocity.
However, the correlation among the plates motion, generated vortices, and aerodynamic forces is still not fully understood yet, especially in the case of multiple bodies.
The DNS (Direct Numerical Simulation, Dong et al., 2006) of freely falling plates is conducted using our in-house CFD (Computational Fluid Dynamics) solver. 
Showcased examples in this video are a part of DNS results in the attempt to answer the above question.

\end{abstract}

\hspace{-0.5cm}\textbf{Reference}

\small
\begin{enumerate}
\item
Willmarth, W. W. and Hawk, N. E. and Harvey, R. L.,  ``Steady and unsteady motions and wakes of freely falling disks'',
Phys. of Fluids, 1964, Vol. 7, pp. 187-208.
\item
Aref H. and Jones, S. W. ``Chaotic motion of a solid through ideal fluid'',
Phys. Fluids A, 1993, pp. 3026-3028.
\item
Belmonte, A. and Elsenberg, H. and Moses E., ``From Flutter to Tumble: Inertial Drag and Froude Similarity in Falling Paper'',
Physical Review Letters, Vol. 81, pp. 345-348.
\item
Andersen, A. and Pesavento U. and Wang, Z.J., ``Unsteady aerodynamics of fluttering and tumbling plates'',
J. of Fluid Mech., 2005, Vol. 541, pp. 65-90.
\item
Andersen, A. and Pesavento U. and Wang, Z.J., ``Analysis of transitions between fluttering, tumbling and steady descent of falling cards'',
J. of Fluid Mech., 2005, Vol. 541, pp. 91-104.
\item
Dong, H., Mittal, R. and Najjar, F.M., ``Wake topology and hydrodynamic performance of low-aspect-ratio flapping foils'',
J. of Fluid Mech, 2006, Vol. 566, pp. 309-343.

\end{enumerate}

\end{document}